# Streaming Multimedia Information Using the Features of the DVB-S Card

Radu Arsinte, Eugen Lupu[1]

**Abstract** – This paper presents a study of audio-video streaming using the additional possibilities of a DVB-S card. The board used for experiments (Technisat SkyStar 2) is one of the most frequently used cards for this purpose.
Using the main blocks of the board's software support it is possible the implement a really useful and full functional system for audio-video streaming. The streaming is possible to be implemented either for decoded MPEG stream or for transport stream. In this last case it is possible to view not only a program, but any program from the same multiplex. This allows us to implement a full functional system useful for educational purposes.

## I. INTRODUCTION

Many consumers are currently using analog TV cards to watch TV on a PC screen. This represents a feature that has already combined the TV and PC experiences. The market for PC analog TV cards has grown over the last few years. With its viewing experience and enhanced interactive data services, a DVB-PC card is a more compelling solution than current analog TV tuner cards for PCs. The DVB experience is significantly enhanced compared with analog TV, as the entertainment is combined with the real-time interactivity of the PC. A potential analog TV card user is inclined to spend a little more for a combination card that supports both analog and DTV broadcasts in order to avoid the risk of quick obsolescence.

Broadcasters will be encouraged to create more DTV content if the installed base of DTVPC cards grows, which will help the DTV industry as a whole.

The main facility for DVB-S ([1]) card users is the possibility to combine local audio-video viewing with streaming. In this way it is possible to give access to audio-video stream for other clients from the same network.

Unlike the Internet, broadcast networks have been optimized for the transmission of rich content to large numbers of users in a predictable, reliable, and scalable manner. The advantages they bring to the infrastructure are as follows:

• Broadcast networks are designed to carry rich, multimedia content. Traditional broadcast networks— including television, cable, and satellite networks- have been explicitly designed to deliver high-quality, synchronized audio and video content to a large population of listeners and viewers. In addition, over the last years many of these networks have migrated from analog to digital transmission systems, thus greatly enhancing their ability to carry new types of digital content, including Internet content.

• Broadcast networks are inherently scalable. By virtue of their point-to-multipoint transmission capability, broadcast networks are inherently scalable. It takes no more resources, bandwidth or other provisions, to send content to a million locations as it does to one, as long all of the receiving locations are within the transmission footprint of the broadcast network. In contrast, with the traditional Internet, each location that is targeted to receive the content will add to the overall resources required to complete the transmission.

• Broadcast networks offer predictable performance. Again, by virtue of the point-to-multipoint nature of transmission on broadcast networks, there are no variances in the propagation delay of data throughout the network, regardless of where a receiver is located. This inherent capability assures a uniform experience to all users within the broadcast network.

Our experience in TS generation ([2],[3]) and use was extremely useful in this work.

## II. DVB-S CARD ARCHITECTURE

Hardware Architecture
The DVB card consists of several components, both on the hardware and software aspects. The basic building block as shown in Figure 3 serves as a platform for standard-definition (SDTV, resolution of 720*576, which is defined as MP@ML (Main Profile at Mail Level)) television program decoding. The main schematic is close to the stand-alone version presented in [4].

---

[1] Facultatea de Electronica, Telecomunicaţii şi Tehnologia Informaţiei, Catedra Comunicaţii, Str. Gh. Bariţiu, 26-28, Cluj-Napoca, email: Radu.Arsinte@com.utcluj.ro

A DVB PC card is composed form a channel-decoding module and a source-decoding module. The channel-decoding module deals with the transmission over the physical media and its main task is to deliver an error free signal to the source-decoding module. It is usually grouped under the term "forward error correction" (FEC) as it provides error detection and correction to the received signal.

On the other hand, the source decoding module descramble, demultiplex and decode the audio and video signal for reproduction. The main task of each functional module is presented in figure 3. The main tasks of source decoding, in a DVB PC-Card, are performed by the host processor (usually at least from Pentium III class).

We are briefly describing the functions of each module:

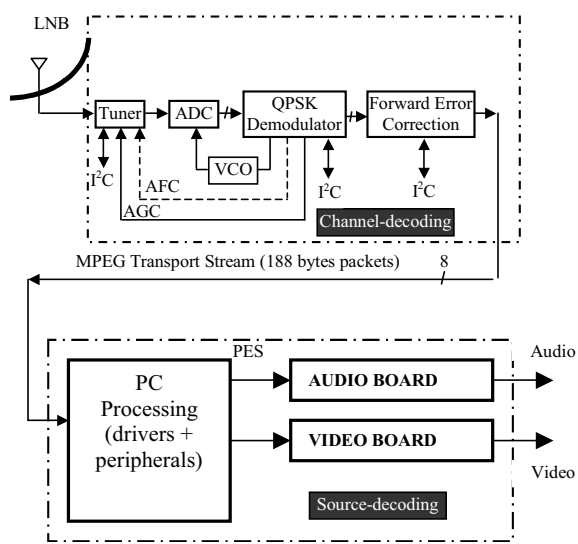

Figure 1. The Hardware Architecture of the DVB Card

1. Tuner

The tuner (sometimes known as the 'front-end'), generally select one of the RF (Radio Frequency) channel and converts it into IF (Intermediate Frequency).

2. ADC (Analogue to Digital Converter)

The ADC receives the analogue signals and converts it into a digital signal for QPSK processing.

3. QPSK Demodulation

This is the key element in the channel decoding process: It performs digital demodulation and half-Nyquist filtering, and reformatting/demapping into an appropriate form for the FEC circuit. It also plays a part in the clock and carrier recovery loops, as well as generating the AGC (automatic gain control) for control of the IF and RF amplifiers at the front end.

4. FFT (Fast Fourier Transform) processor

The FFT processor provides timing and frequency synchronization, channel estimation and equalization, generation of optimal soft decisions using the channel state information, symbol and bit de-interleaving.

5. Forward Error Correction (FEC)

The FEC block performs de-interleaving, Reed-Solomon decoding and energy dispersal de-randomizing. The output data are the 188 bytes transport packets in parallel form (8 bit data, clock and control signals). The channel-decoding module is highly integrated and it is usually offered as a single chip solution (module 2 to 5).

Software Architecture

The software required to power the DVB set-top boxes or PC cards is apparently more complex than the hardware requirement since most of the hardware are already highly integrated. An example of the software model used for the development is presented in figure 2.

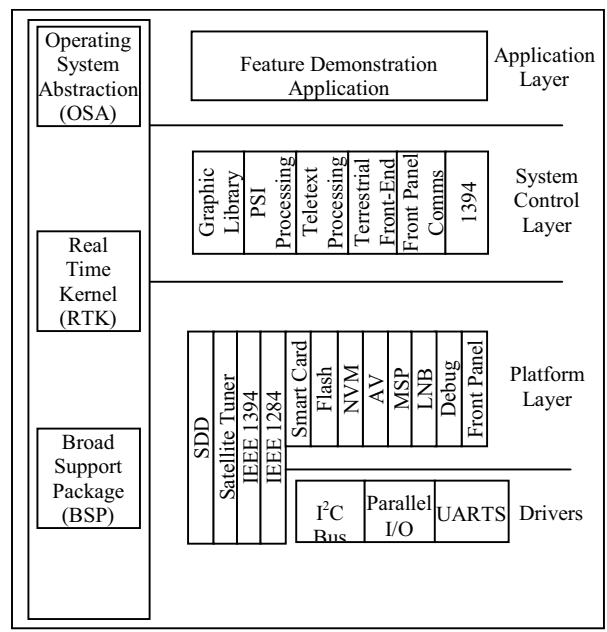

Figure 2: Software Architecture for the DVB-S Card

The task of decoding is mainly done in hardware, the software deals with configuring these devices upon power up and to handle user requests. Most of the software modules are required to program the EPG (Electronic Program Guide) and Interactive TV (if included). Not all the components are required for a specific function (TV reception, for example), but the software driver could be updated permanently to match new requirements in functionality. Our experience in TS generation and use was extremely useful in this work.

III. STREAMING OPTIONS

Streaming information in a DVB-S based environment, is not different basically from a normal network-based information streaming. The main differences are related with a minimal bandwidth necessary to transfer al large amount of information, characteristic to multimedia information. A rough estimation of the compression/storage ratio obtained by a common MPEG2 encoding is: 1.8Gbytes per

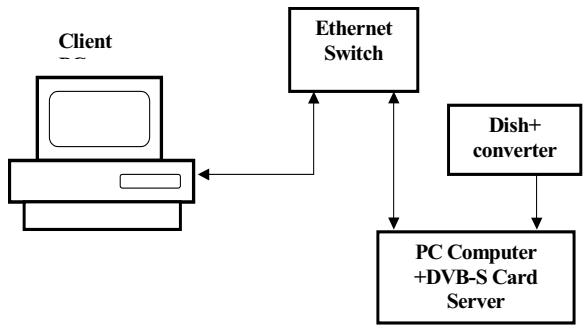

Figure 3.a Test connection using a switch

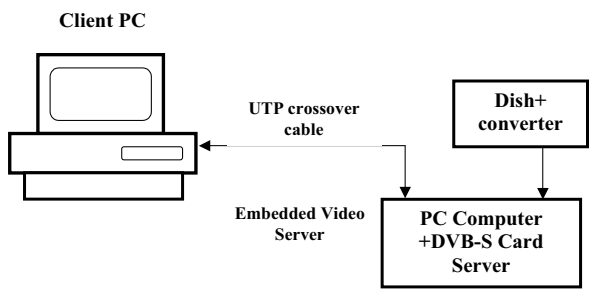

Figure 3.b Test connection using a direct cable

hour, for instance to store a full screen DVD in average quality on your harddisk. In terms of real-time streaming, this means about 4 Megabits per second (4Mb/s), which can also be quantified as 582 KiloBytes per second (582KB/s).
Here is a short explanation of the above calculations:
*1. 2GigaBytes of stored video /60 mins /60 secs = 596523 bytes per sec*
   *a. 596523 bytes per sec /1024 = 582 KB/s (Kilobytes per sec)*
*2. 596523 bytes per sec *8 = 4772184 bits per sec*
   *a. 4772184 bits per sec /1024 /1024 = 4 Mb/s (Megabits per sec)*

Hardware configuration
The test server uses basic functions of the Video server.
The minimal configuration of the test system is composed from a server (PC system with DVB-S board) and a client (a normal PC). The proposed configurations are presented in Fig.3 (a and b), similar with the configurations presented in ([5]) or ([6]).
If this test system (involving a local network and a DHCP server – Fig.3.a) is not possible to be implemented, it is possible to use a simplified version based on a direct connection using a UTP crossover cable (Fig.3.b).

Software configuration
The main element for streaming is the Server4PC utility (described also in [7]) delivered with SkyStar board. This utility must be configured properly to perform the requested tasks.

It is necessary to provide the IP-address of the network interface., the multicast stream is sent for for distribution. A common address used in most test applications is 192.168.0.1.
The second information, necessary for IP multicast is the multicast address and the port, where the stream is located. The multicast IP range is specified in RFC1112. Multicast IP address are defined to the range between 224.0.0.0 to 239.255.255.255. To send the stream to all clients from the subnet, it is necessary to use the multicast IP 224.0.0.1. The multicast port number, can be chosen in the range 0 and 65500. The first 1024 ports are reserved for IP services.
The main settings are presented in a screen capture from figure 4.

The server part uses a well-known program called VLC media Player, developed in VideoLAN project. The main screen of the program is presented in figure 5.

The VideoLAN project targets multimedia streaming of MPEG-1, MPEG-2, MPEG-4 and DivX files, DVDs, digital satellite channels, digital terrestial television channels and live videos on a high-bandwidth IPv4 or IPv6 network in unicast or multicast under many OSes. VideoLAN also features a cross-plaform multimedia player, VLC, which can be used to read the stream from the network or display video read locally on the computer under all GNU/Linux flavours, all BSD versions, Windows, Mac OS X, BeOS, Solaris, QNX, Familiar Linux.

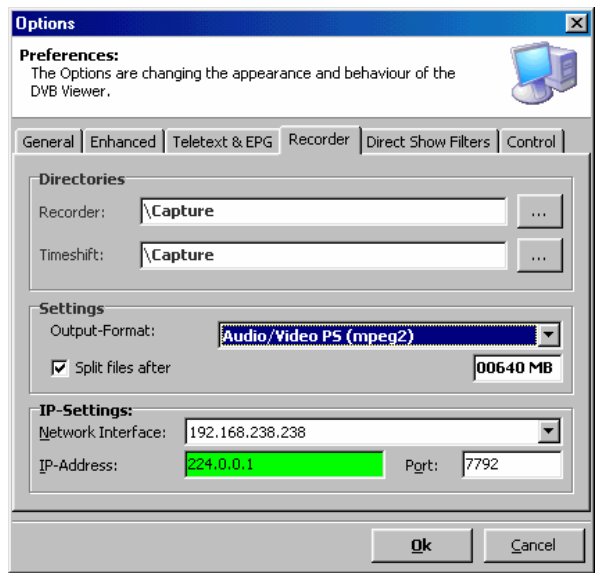

Figure 4. DVB Viewer configuration

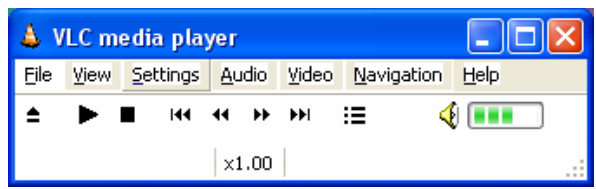

Figure 5. Main screen of VLC Media Player

The latter feature is one of the most important points for VideoLan, being in fact the only free software being able to fill the gaps between video and audio formats on different platforms, while it's even supporting ARM and MIPS based hand-held devices, being ready to be integrated in consumer-grade embedded solutions. VideoLan supports the functionality offered by the IVTV driver, taking advantage of hardware encoding on a growing number of cards. It is basically made out of two software components: VideoLanClient (VLC) and VideoLanServer (VLS). VLC is a multipurpose streaming a/v client and source: it can play streams as well it can capture and send a stream to another VLC or VLS. VLS is just a server which has no visualization output for the streams it handles, its fully capable of capturing and streaming from the local machine, as well reflect streams coming from other VLC/VLS nodes. The way VideoLan distributes functionality among its nodes is therefore very flexible and permits to easily build streaming topologies to distribute real-time audio/video streams.

## IV. RESULTS

The proposed configuration was implemented in a local network with 5 computers – one as server and four clients. The performance of the computers is the following:
Server: Pentium II – 400MHz -10GB- 128MB RAM
Clients: Pentium II-III – 233...800MHz - 128-256MB RAM
Experiments proved that the streaming is possible for all the computers, with some limitations for low-grade computers (Pentium II – 233).

The goal of the work being an investigation of this technology for an educational use, the results could be interpreted as satisfactory.

We tried to extend the streaming over a wider network, this fact being extremely difficult without modifying the network security settings.

## V. CONCLUSION AND FUTURE WORK

Our activity brought us the following achievements:
1. Installing and configuring the driver of DVB-S card, eliminating any incompatibilities;
2. Streaming of information in a local network;
3. Configuring the server for the PC containing the DVB card;
4. Configuring the client applications using both the board's viewer and VLC;
5. Measurements of the streaming performance in the network.

The experiments revealed that an efficient streaming is possible to be implemented, but the quality of the network is essential. The uninterrupted streaming (and of course viewing) could be realized only in moderate loaded networks.

All the test work was done in a Windows environment (both for client or server). The next experiments will try to verify and measure the same performance in a Linux based environment, or a hybrid environment (for example Linux for server, and Windows for clients).

We will try also to expand the test network, to be able the send Transport Streams in a large area (for example in entire faculty network).